\documentclass[nojss]{jss}

\usepackage{orcidlink,thumbpdf,lmodern}

\usepackage{framed}

\author{Skerdi Haviari~\orcidlink{0009-0008-1104-3037}
\\Université Paris Cité
\\ Assistance Publique - Hôpitaux de Paris
               }

\title{Adaptive Sample Size Simulations with \proglang{R} package \pkg{adsasi}}
\Plaintitle{Adaptive Sample Size Simulations with R package adsasi}
\Shorttitle{Adaptive Sample Size Simulations with \pkg{adsasi}}

\Abstract{
Planning empirical experiments such as clinical trials or A/B tests requires sample size determination, which in many interesting cases has no closed-form solution (e.g. factorial or adaptive designs). \pkg{adsasi} is a new \proglang{R} package that enables simulations-first sample size calculations for any trial that can be simulated in short compute time. First, the user specifies as a function that takes a sample size as argument, simulates the experiment, and returns a boolean for success/failure. Then, \pkg{adsasi} functions \code{adsasi\_0d} and \code{adsasi\_1d} iteratively call it on different sample sizes and progressively home in on the one with nominal success rate (power), assuming that increasing sample size increases power. \code{adsasi\_1d} can also draw, purely empirically, the relationship between a design parameter and sample size. The implementation uses a modified probit regression (with success/failure as the dependent variable), informed by simulations conducted around the target size, and provides standard errors at each stage using the Cramér-Rao bound derived from a custom analytical Hessian matrix. Simple examples are first presented, yielding results within Monte Carlo variance of their closed-form expressions, then intractable ones (including bootstrapping from an existing medical cohort). \pkg{adsasi} will hopefully facilitate the funding and conduct of interesting, highly complex experimental designs by making their sizing straightforward.
}

\Plainkeywords{clinical trial simulation, sample size, R}

\Address{
  Skerdi Haviari\\
  Université Paris Cité and Université Sorbonne Paris Nord\\
  Inserm, IAME\\
  F-75018 Paris, France\\
  \emph{and}\\
  Département Epidémiologie, Biostatistiques, Recherche Clinique\\
   AP-HP, Hôpital Bichat Claude Bernard\\
  F-75018 Paris, France\\
  E-mail: \email{skerdi.haviari@aphp.fr}\\
}

\begin{document}

%% -- Introduction -------------------------------------------------------------

%% - In principle "as usual".
%% - But should typically have some discussion of both _software_ and _methods_.
%% - Use \proglang{}, \pkg{}, and \code{} markup throughout the manuscript.
%% - If such markup is in (sub)section titles, a plain text version has to be
%%   added as well.
%% - All software mentioned should be properly \cite-d.
%% - All abbreviations should be introduced.
%% - Unless the expansions of abbreviations are proper names (like "Journal
%%   of Statistical Software" above) they should be in sentence case (like
%%   "generalized linear models" below).

\section{Introduction} \label{sec:intro}

Calculation of sample size is often a cornerstone of the design and practical implementation of a clinical trial, and any high-variability experiment more generally. Many closed-form formulae are available, often relying on asymptotic approximations. However, in everyday clinical trial practice, particular circumstances that eschew closed-form solutions will often arise, such as particular patterns of drop-outs, adaptive randomization, adaptive changes in the analysis model (e.g. depending on interactions), a need to hit multiple endpoints simultaneously, or expected changes in standard of care. In this context, practitioners will often turn to simulations, but those are more often used at the verification stage (checking that the approximated size provided adequate alpha and beta risk) rather than the derivation stage. 

Building on a tool that we used for a previous unusual trial design study  \citep{haviari_distributive_2024}, we have created the \pkg{adsasi} package for \proglang{R}, to enable a simulations-first approach to sample size computations, completely obviating the need for sample size formulae, unless the inference itself is time-consuming. 

Meant for users that are comfortable with writing thir own functions in \proglang{R}, \pkg{adsasi} requires the user to write a simulator for a single trial, that can be as complex as their situation requires, and then directly pass it to the \pkg{adsasi} wrappers to automatically explore different sample sizes and find the desired one. One of the two functions can also find an empirical optimum for a variable design parameter. 

\section[Other R packages and methods]{Other \proglang{R} packages and methods} \label{sec:review}

Four clinical trial simulation packages in \proglang{R} are widely known : \pkg{Mediana} \citep{paux_mediana_2019}, \pkg{SimEngine} \citep{kenny_simengine_2024}, \pkg{TrialSimulator} \citep{zhang_trialsimulator_2026}, and \pkg{simtrial} \citep{anderson_simtrial_2025}. These packages are mostly used to write simulation functions of the kind that \pkg{adsasi} takes as input, but have no specific tools to try different sample sizes adaptively and empirically : for example, a vignette from \pkg{SimEngine} on sample size by simulation advises the user to ``guess and check'' \citep{kenny_example_2025}. When faced with sample size problems, these packages would likely benefit from being integrated with \pkg{adsasi}, which is specifically designed to do the ``guess and check'' part automatically and reliably, somewhat like a human would, but with more accuracy and stamina. 

 The package \pkg{nRegression} \citep{shilane_nregression_2023} also contains one function that tries different sample sizes, but without smooth information sharing, in contrast with \pkg{adsasi}. It is also aimed at regression problems, that can be addressed by another popular package, \pkg{PFIM} \citep{leroux_pfim_2026}. 
\pkg{PFIM} can find sample sizes and perform optimization using the Fisher Information Matrix (the inverse of the covariance matrix), to optimize accuracy for the estimation of different parameters. For problems that have a clear-cut covariance matrix, PFIM will generally be a near-optimal approach, because the covariance matrix has all the available information. However, there will be many cases where such a clear-cut covariance matrix does not exist, typically when the model is selected based on the data, including when the experiment itself is adaptive, when the output is not a model coefficient at all (but rather a ranking, or a binary condition), or when multiple endpoints have to be reached simultaneously. For this general type of problem, specific solutions are not easily found. 

On the algorithmic side, few studies seem to have been done on this specific topic. It is possible to view the problem as a special case of optimization of a stochastic objective function, for which widely used algorithms such as Adam \citep{kingma_adam_2017} are available, but there are several specificities to the problem of sample size that would require significant work from the user. First, trial designers want to describe a tradeoff and not just find an optimum (the sample size for maximum statistical power is trivial, it is infinity), which would require a specific formulation of the objective function. Second, there are specific relationships to this problem that should be coded in the algorithm for maximum efficiency (e.g., that increasing sample size cannot decrease power). Third, simulations with a binary outcome are a lot more noisy than commonly envisioned stochastic objective functions, likely making such approaches suboptimal without adaptation to this context. 

Outside of stochastic objective optimization, \cite{wilson_efficient_2021} provide a method for generalized sample size computations by simulation, but offer an expensive approach that often simulates above the desired sample size, and in relatively large batches outside the region of interest. In particular, it initiates with 30-50\% of the budgeted computation, most of which will be wasted unless the user already has a very good intuition. The method also does not explicitly model the relationship between design parameters, sample size and power, likely reducing its efficiency. Finally, the algorithm is not, to date, implemented in an easily imported package. 

%% -- Manuscript ---------------------------------------------------------------

%% - In principle "as usual" again.
%% - When using equations (e.g., {equation}, {eqnarray}, {align}, etc.
%%   avoid empty lines before and after the equation (which would signal a new
%%   paragraph.
%% - When describing longer chunks of code that are _not_ meant for execution
%%   (e.g., a function synopsis or list of arguments), the environment {Code}
%%   is recommended. Alternatively, a plain {verbatim} can also be used.
%%   (For executed code see the next section.)

%% -- Manuscript ---------------------------------------------------------------

%% - In principle "as usual" again.
%% - When using equations (e.g., {equation}, {eqnarray}, {align}, etc.
%%   avoid empty lines before and after the equation (which would signal a new
%%   paragraph.
%% - When describing longer chunks of code that are _not_ meant for execution
%%   (e.g., a function synopsis or list of arguments), the environment {Code}
%%   is recommended. Alternatively, a plain {verbatim} can also be used.
%%   (For executed code see the next section.)

\section{Methods} \label{sec:methods}
\subsection{Intuition and simulation-wise likelihood} \label{sec:intuition}

The basic intuition of the \pkg{adsasi} simulation wrappers is that for sample size problems under the alternate hypothesis with an unbiased estimator, low sample sizes will have low power (either 0 or the $\alpha$ risk), and as sample size increases, the power will cross the desired value and then keep increasing up to almost 100\% for very large sample sizes. This monotonicity can be leveraged to share simulation information across close sample sizes, and then to pick new ones to try. If one knows the true value is close, a simple linear equation could model power well enough, but for simulations that are computed further from the desired value, the approximation will break down. Hence the need for a more robust link. 

Many equations could model this relationship, but for simplicity and derivability, we use the probit function. This is congruent with the shape of most closed-form sample size formulae
\begin{equation} \label{eq:1}
N={(z_{1-\alpha/2}+z_{1-\beta})}^2/K
\end{equation}
where $K$ is an expression of the effect size incorporating variability, $z$ denotes the standard normal reciprocal cumulative distribution function, and $\alpha$ and $\beta$ are respectively type I and II error. Rearranging Equation~\ref{eq:1} to express $1-\beta$ as a function of $N$, we get 
\begin{equation} \label{eq:2}
1-\beta=\Phi(a+b\sqrt{N})
\end{equation}
where $1-\beta$ is the probability that a simulated trial is positive (amenable to binomial regression with a well-defined likelihood), $\Phi$ is the standard normal cumulative distribution function (CDF) and $a$ and $b$ are functions of $\alpha$ and $K$. Although we derive the function from the simple case of a single statistical test, we use the terms ``positive'' or ``successful'' rather than ``significant'' to emphasize that the user may not be aiming for a certain power on a single test, but rather for a certain probability of a more complex favorable outcome (such as several significant endpoints, or a true ranking of more than 2 interventions). 

We further reparametrize this linear function of $\sqrt{N}$ as
\begin{equation} \label{eq:3}
1-\beta=\Phi\left(z_{1-\beta_0}+e^s(X-X_0)\right)
\end{equation}
where $X$ is the square root of the sample size, $X_0$ is the square root of the (unknown) sample size that gives a power of $1-\beta_0$ (the target power), $e^s$ is a (positive) slope that characterizes how fast power deviates from $1-\beta_0$ as $X$ deviates from $X_0$, and $z_{1-\beta_0}$ ensures that the required power is obtained when $X=X_0$. This formula has two unknown parameters $s$ and $X_0$ and encodes a well-defined likelihood, allowing maximum-likelihood estimation from the data obtained by drawing simulations at different values of $X$ (i.e., different sample sizes). 

If the design can vary according to a particular parameter $v$, we can further posit that $s$ and $X_0$ will be functions of $v$. $v$ could be an allocation ratio, a follow-up time for time-to-event outcomes, etc. In complex cases that are of most interest, the equation linking them will not be known, and an assumption will have to be made. For simplicity, since finding exact but very complex relationships is generally not the objective when taking an empirical approach, we will use polynomials of the form : 
\begin{equation} \label{eq:4}
s\left(v\right)=s_0+s_1v^1+s_2v^2+s_3v^3+\ldots
\end{equation}
\begin{equation} \label{eq:5}
X_0\left(v\right)=x_0+x_1v^1+x_2v^2+x_3v^3+\ldots
\end{equation}
The exact number of polynomial terms will be left to the user, with default values empirically found to work best for most problems, and some advice for unusual cases (i.e., the more complex the expected relationship, the more terms). 

\subsection{Partial likelihood, gradient and Hessian for the fixed design case} \label{sec:eq0d}

When the design is fixed, $s$ and $X_0$ are not functions of a varying parameter, so the partial likelihood $\mathcal{L}_i$ of a given simulation $i$ of size $N_i$ with outcome $O_i$ is either (by application of Equation~\ref{eq:3})
\begin{equation} \label{eq:6}
\mathcal{L}_i\left(s,X_0\middle| O_i=1\right)=\Phi\left(z_{1-\beta_0}+e^s(\sqrt{N_i}-X_0)\right)
\end{equation}
if the trial has a positive outcome, or 
\begin{equation} \label{eq:7}
\mathcal{L}_i\left(s,X_0\middle| O_i=0\right)=1-\Phi\left(z_{1-\beta_0}+e^s(\sqrt{N_i}-X_0)\right)
\end{equation}
if it does not. For simplicity, we write
\begin{equation} \label{eq:8}
\mathcal{L}_i\left(s,X_0\right)=\Phi_\pm\left(I_i\right)
\end{equation}
where conditionality on outcome is implied and $\Phi_\pm$ denotes either the lower section (positive outcome) or the upper section (negative outcome) of the normal CDF, depending on $O_i$, and $I_i$ (``inside'') denotes $z_{1-\beta_0}+e^s(\sqrt{N_i}-X_0)$. 

The derivative of the partial log-likelihood of trial $i$ with respect to $X_0$ is 
\begin{equation} \label{eq:9}
\frac{d\ln\mathcal{L}_i\left(s,X_0\right)}{dX_0}=\frac{d\ln\left(\Phi_\pm\left(I_i\right)\right)}{dX_0}=\frac{dI_i}{dX_0}\frac{d\ln\left(\Phi_\pm\left(I_i\right)\right)}{dI_i}=-e^s\frac{\pm\varphi\left(I_i\right)}{\Phi_\pm\left(I_i\right)}
\end{equation}
where $\pm$ denotes +1 for successful simulated trials and -1 for unsuccessful ones (congruent with $\Phi_\pm$ above) and $\varphi$ is the standard normal probability distribution function (PDF). Similarly, the derivative with respect to $s$ is 
\begin{equation} \label{eq:10}
\frac{d\ln\mathcal{L}_i\left(s,X_0\right)}{ds}=\frac{dI_i}{ds}\frac{d\ln\left(\Phi_\pm\left(I_i\right)\right)}{dI_i}=S_i\frac{\pm\varphi\left(I_i\right)}{\Phi_\pm\left(I_i\right)}
\end{equation}
where $S_i$ (``shift'', for the \emph{power} shift in probit space) denotes $e^s\times(\sqrt{N_i}-X_0)$ or (they are equal) $I_i-z_{1-\beta_0}$. These values constitute the gradient that is used for the two-parameter optimization in the fixed-design case. 

For the Hessian, we pre-compute the general term 
\begin{equation} \label{eq:11}
\frac{d}{du}\left(\frac{\ \varphi\left(I_i\right)}{\Phi_\pm\left(I_i\right)}\right)=\frac{dI_i}{du}\frac{\varphi\prime\left(I_i\right)\Phi_\pm\left(I_i\right)-\pm\varphi^2\left(I_i\right)}{{\Phi_\pm}^2\left(I_i\right)}=\frac{dI_i}{du}D_i
\end{equation}
where $u$ can be any variable ($s$ or $X_0$), $\varphi\prime$ is the derivative of the standard normal PDF, and $D_i=\frac{\varphi\prime\left(I_i\right)\Phi_\pm\left(I_i\right)-\pm\varphi^2\left(I_i\right)}{{\Phi_\pm}^2\left(I_i\right)}$ will be a recurring expression in our derivations. We use this equation to compute the $2\times2$ Hessian, getting 
\begin{equation} \label{eq:12}
\frac{d\left(\ln\mathcal{L}_i\left(s,X_0\right)\right)^2}{ds^2}=\pm S_i\left(\frac{\varphi\left(I_i\right)}{\Phi_\pm\left(I_i\right)}+S_iD_i\right)
\end{equation}
\begin{equation} \label{eq:13}
\frac{d\left(\ln\mathcal{L}_i\left(s,X_0\right)\right)^2}{dX_0ds}=-\pm e^s\left(\frac{\varphi\left(I_i\right)}{\Phi_\pm\left(I_i\right)}+S_iD_i\right)
\end{equation}
\begin{equation} \label{eq:14}
\frac{d\left(\ln\mathcal{L}_i\left(s,X_0\right)\right)^2}{d{X_0}^2}=\pm e^{2s}D_i
\end{equation}
These values are then implemented in the size-finding algorithm itself (see Section~\ref{sec:algo}) to obtain standard errors under the assumption that the Cramér-Rao bound is reached (which is reasonable with very local sampling and a high number of simulations). 

\subsection{Partial likelihood, gradient and Hessian for the varying design case} \label{sec:eq1d}

Using similar notations as above and plugging in the polynomials from Equations~\ref{eq:4} and~\ref{eq:5}, we have 
\begin{equation} \label{eq:15}
\mathcal{L}_i\left(s_0,s_1,s_2,s_3,\ldots x_0,x_1,x_2,x_3,\ldots \right)=\Phi_\pm\left(I_i\right)
\end{equation}
where $I_i$ now denotes $z_{1-\beta_0}+e^{s_0+s_1v_i^1+s_2v_i^2+\ldots }\left(\sqrt{N_i}-\left(x_0+x_1v_i^1+x_2v_i^2+\ldots \right)\right)$, $v_i$ being the varying parameter value for simulation number $i$. For brevity we now also write $S_i = e^{s\left(v_i\right)}\left(\sqrt{N_i}-X_0\left(v_i\right)\right)$ (same \emph{power shift} in probit space as above). $D_i$ also keeps the same meaning as in Equation~\ref{eq:11}, extended to the varying parameter setting. 

For the gradient, for each coefficient $s_0,s_1,s_2,\ldots$ (indexed by subscript $j$) and $x_0,x_1,x_2,\ldots$ (indexed by subscript $k$), abbreviating the partial likelihood as $\mathcal{L}_i\left(s_0,\ldots x_0,\ldots \right)$, we have respectively
\begin{equation} \label{eq:16}
\frac{d\ln\left(\mathcal{L}_i\left(s_0,\ldots x_0,\ldots \right)\right)}{ds_j}=\frac{d\left(I_i\right)}{ds_j}\frac{\pm \varphi\left(I_i\right)}{\Phi_\pm\left(I_i\right)}=\pm v^jS_i\frac{\ \varphi\left(I_i\right)}{\Phi_\pm\left(I_i\right)}
\end{equation}
and
\begin{equation} \label{eq:17}
\frac{dln\left(\mathcal{L}_i\left(s_0,\ldots x_0,\ldots \right)\right)}{dx_k}=\frac{d\left(I_i\right)}{dx_k}\frac{\pm \varphi\left(I_i\right)}{\Phi_\pm\left(I_i\right)}=-\pm v^ke^{s\left(v_i\right)}\frac{\ \varphi\left(I_i\right)}{\Phi_\pm\left(I_i\right)}
\end{equation}

Deriving the Hessian for each simulated trial by blocks, with the help of Equations~\ref{eq:11},~\ref{eq:16} and~\ref{eq:17}, using $j\prime$ and $k\prime$ when parameters from the same polynomial are involved, we have : 
\begin{equation} \label{eq:18}
\frac{d\left(\ln\mathcal{L}_i\left(s_0,\ldots x_0,\ldots\right)\right)^2}{ds_jds_{j\prime}}=\pm v^jv^{j\prime}S_i\left(\frac{\ \varphi\left(I_i\right)}{\Phi_\pm\left(I_i\right)}+S_iD_i\right)
\end{equation}
\begin{equation} \label{eq:19}
\frac{d\left(\ln\mathcal{L}_i\left(s_0,\ldots x_0,\ldots\right)\right)^2}{ds_jdx_k}\ =-\pm v^jv^ke^{s\left(v\right)}\left(\frac{\ \varphi\left(I_i\right)}{\Phi_\pm\left(I_i\right)}+S_iD_i\right)
\end{equation}
\begin{equation} \label{eq:20}
\frac{d\left(\ln\mathcal{L}_i\left(s_0,\ldots x_0,\ldots\right)\right)^2}{dx_kdx_{k\prime}}=\pm v^kv^{k\prime}e^{2s\left(v\right)}D_i
\end{equation}
which are, as intuitively expected, generalizations of Equations~\ref{eq:12},~\ref{eq:13} and~\ref{eq:14}. These partial Hessian terms allow us to compute the simulations set-wide Hessian which is used to obtain standard errors for any value of $X_0\left(v\right)$, again under the Cramér-Rao approximation. 

\subsection{Algorithm} \label{sec:algo}
\pkg{adsasi} functions require the user to input a function that they have created, which takes a sample size as argument and returns a Boolean indicating success (desired outcome) or failure of the trial. 

For the simple case implemented in \code{adsasi_0d} (\emph{zero-dimensional}, since the design is fixed except for sample size), the function first draws 50 sample sizes (between 4 and a user-provided upper limit) at which to simulate trials, simulates them using the user-provided function, and fits the probit model above to find the values of parameters $s$ and $X_0$. It then finds the standard error ($SE$) of $X_0$ using the inverse of the Hessian and draws the next batch of simulations uniformly in the $X_0\pm1$SE window. The process is repeated, first in batches of 50 until 500 simulations are done, then in batches of 10\% of existing simulations, until the wanted number of simulations is reached. In practice, the algorithm quickly ``zooms in'' on the right sample size and progressively starts basically repeatedly checking the right value (and values close to it). 

For the optimizing case implemented in \code{adsasi_1d} (\emph{one-dimensional}, since an optimization over one dimension will occur), the idea is the same, except the initiation is made over a wider space. The tunable parameter $v$ is chosen between -1 and 1 (so that the polynomials behave in a regular way) and then scaled to the window given by the user before being passed to the simulation function they provided (so these user-provided arguments, namely the simulation function and the optimization window, have to be congruent). Using estimates for the $x_k$ terms, estimations of $X_0\left(v\right)$ (the quare root of the sample size) for 201 values of $v$ between -1 and 1 are computed, explicitly deriving a relationship between $v$ and sample size. The log-likelihood Hessian is also used to compute the Cramér-Rao bound, that is assumed to be reached given the high number of simulations. Whereas this yielded the standard error for $X_0$ directly in \code{adsasi_0d}, in \code{adsasi_1d} standard covariance formulae have to be applied to obtain the variance of the polynomials $s\left(v\right)$ and $X_0\left(v\right)$ for each value of $v$. This is straightforward since said polynomials are simply linear combinations of coefficients when holding $v$ constant. The roots of these variances are the standard errors of interest. This straightforward computation of standard errors is the reason why we did not simply use a native probit regression within \proglang{R}. 

For both \code{adsasi_0d} and \code{adsasi_1d}, sample sizes for the next batch of simulations are drawn in a $\pm1$SE window around the estimate, limiting the effect of mis-specifications by staying close to the empirical right value. For \code{adsasi_1d}, values of $v$ for the next batch of simulations are drawn with a probability proportional to the inverse squared predicted sample size at each value of $v$ (i.e., values of $v$ with lower sample size are favored). 

After each batch, the inference for the coefficients is run again and the process is repeated until the wanted number of simulations is reached. 

\section{Accuracy of the sample size finders}\label{sec:acc}
For both \code{adsasi_0d} and \code{adsasi_1d}, typical accuracy is shown visually in this section. For this, we pick a simple design with a known closed-form solution, namely a $t$~test,  on a randomized experiment with equal or unequal allocation between two arms. 
\subsection{Fixed design, $t$~test} \label{sec:zeroacc}
The first wrapper of the package, \code{adsasi_0d}, finds the sample size for a fixed design (hence the suffix 0d, for \emph{zero-dimensional}). The equations are in Section~\ref{sec:eq0d}, the algorithm is described in Section~\ref{sec:algo}. 
The introduction and validation case presented here is that of a simple $t$~test. For a given sample size (that we will call \code{NN}), the user will need to write a simulation function that will become the argument \code{simfun} for \code{adsasi_0d}. In our first example, the function will sample a population of \code{NN} patients, randomize them among two arms with a fraction of (by default) \code{f1 = 0.5} in the first arm, apply a user-specified effect size of (by default) \code{delta = 0.2} standard deviations, run a $t$~test and check whether its $p$~value is lower than the user-specified $\alpha$ risk of (by default) \code{alpha = 0.05}. The syntax is shown below. 
\begin{CodeChunk}
\begin{CodeInput}
R> simulate_unequal_t_test = function(NN = 20, f1 = 0.5, 
+                                     delta = 0.2, alpha = 0.05)
+    {
+    n1 = round(f1 * NN)
+    n2 = NN - n1
+    yy1 = rnorm(n1)
+    yy2 = rnorm(n2, delta)
+    pp = NA ; try(pp <- t.test(yy1, yy2)$p.value, silent = TRUE)
+    !is.na(pp) & pp < alpha
+    }
\end{CodeInput}
\end{CodeChunk}
The function can have any name. Note the robust syntax at the end, using \code{try()} and \code{!is.na()} rather than simply \code{t.test(yy1, yy2)$p.value<alpha}. This is because \code{adsasi_0d} will be trying many values for \code{NN}, including low ones for which a standard $t$~test makes no sense (for example, one of the arms will have a single patient). \code{adsasi_0d} requires the simulation function to never output \code{NA}, so the user has to ensure that. While programming \code{adsasi_0d} to interpret \code{NA} as \code{FALSE} would have been possible, this would have masked potentially severe user errors and was thus intentionnally avoided. 

We can then call this function simply with 
\begin{CodeChunk}
\begin{CodeInput}
R> set.seed(0)
R> adsasi_0d(simulate_unequal_t_test)
\end{CodeInput}
\end{CodeChunk}
and obtain, as output, the sample size for the default 90\% power (graphical outputs are also returned, which are covered in Section~\ref{sec:args}). 
\begin{CodeChunk}
\begin{CodeOutput}
$size_estimate
[1] 1020
\end{CodeOutput}
\end{CodeChunk}
Note that \code{adsasi_0d} will be modifying the argument named \code{NN}, which must be a named argument representing total sample size in the user-supplied function. The other arguments use their default values unless specified, in particular \code{f1 = 0.5} which means the arms have the same size. 

In order to evaluate the accuracy of the algorithm, we replicate the computation 1000 times with \code{nsims = 2000} simulations for each sample size computation. 
\begin{CodeChunk}
\begin{CodeInput}
R> set.seed(0)
R> results_fig1 = unlist(replicate(1000,
+    adsasi_0d(simulate_unequal_t_test, nsims = 2000, capNN = 5000)
+    ))
\end{CodeInput}
\end{CodeChunk}
In the call above, we set a higher ceiling than the default for authorized simulation sample sizes with \code{capNN = 5000} in order to avoid hindering exploration for a minority of outlier runs. 

\begin{figure}[t!]
\centering
\includegraphics{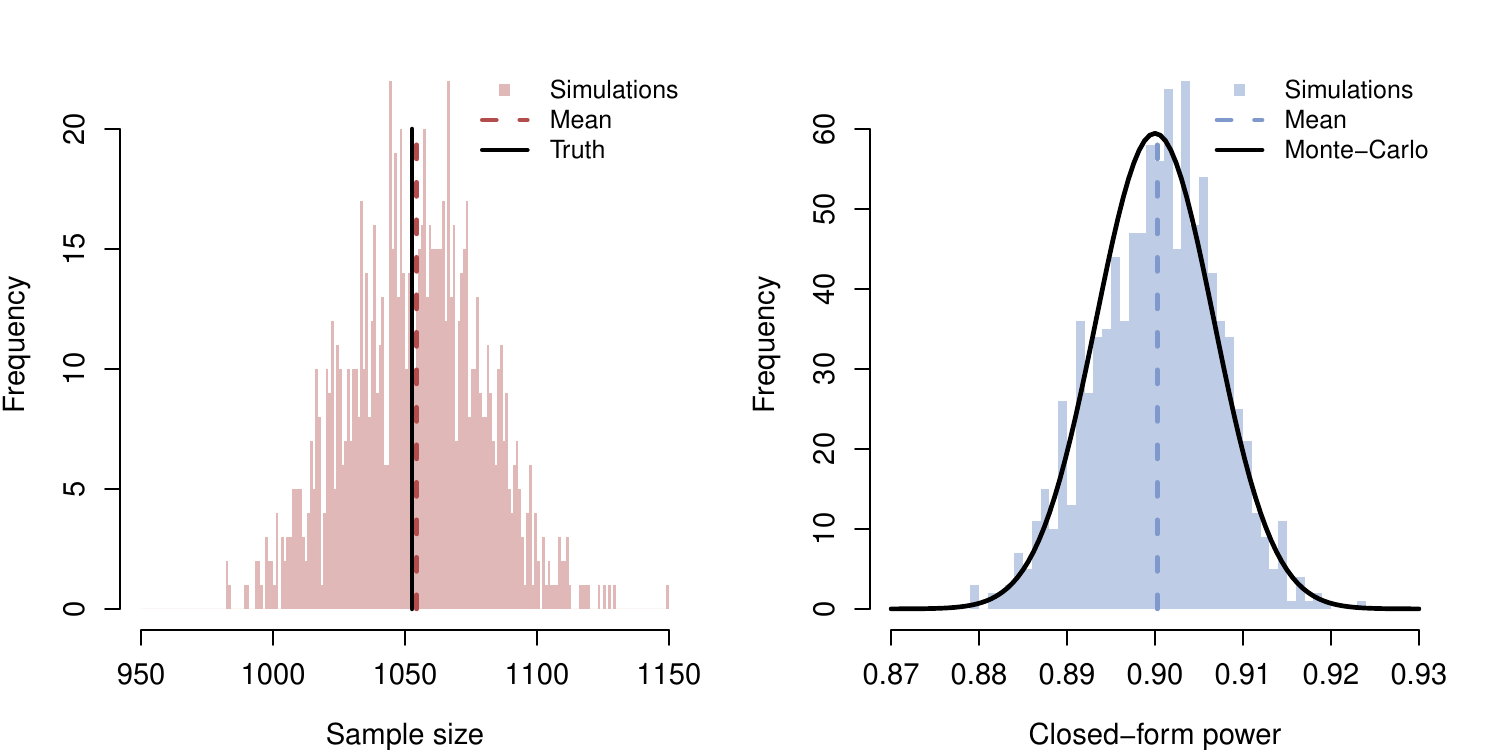}
\caption{\label{fig:fig1} Accuracy of \code{adsasi\_0d}. (left) Empirically found sample sizes versus analytical value. (right) Closed-form power of the returned sample sizes, versus Monte Carlo variability with the same number of draws, in a true proportion of 90\%, as \code{adsasi\_0d} is using to make its inference.}
\end{figure}

We can then plot the 1000 resulting sample sizes found, as shown in Figure~\ref{fig:fig1}. This distribution is compared to the true sample size (left panel), and to the expected variability due to pure Monte Carlo variance (right panel). The latter merits some explanation : because \code{adsasi_0d} is purely empirical and simulates trials around what it estimates to be the correct sample size, we can expect some variability in the power of the sample size it returns. How much ? With 2000 simulations, we expect the variability to be approximately that of a true proportion of 90\% sampled 2000 times, because the 2000 simulations on which \code{adsasi_0d} bases its inference are close to the right value, and have a similar Monte Carlo variance. Because we know the analytical power for a $t$~test, we can compute the value for the 1000 returned sample sizes, and plot them in the right panel of Figure~\ref{fig:fig1}, with a very good overlap with what is expected. 

This Monte Carlo binomial variance is a good rule to decide how many simulations to run. For example, if one wants the returned sample size targeting 90\% power to deviate by no more than 1 percentage point from said target (meaning a 95\% interval of width equal to 2 percentage points), 3500 simulations will be needed. This is simply obtained with the variance of a proportion under the normal approximation $\frac{p\left(1-p\right)}{N}$.

\subsection{Optimizing the design, unequal $t$~test} \label{sec:oneacc}
\begin{figure}[t!]
\centering
\includegraphics{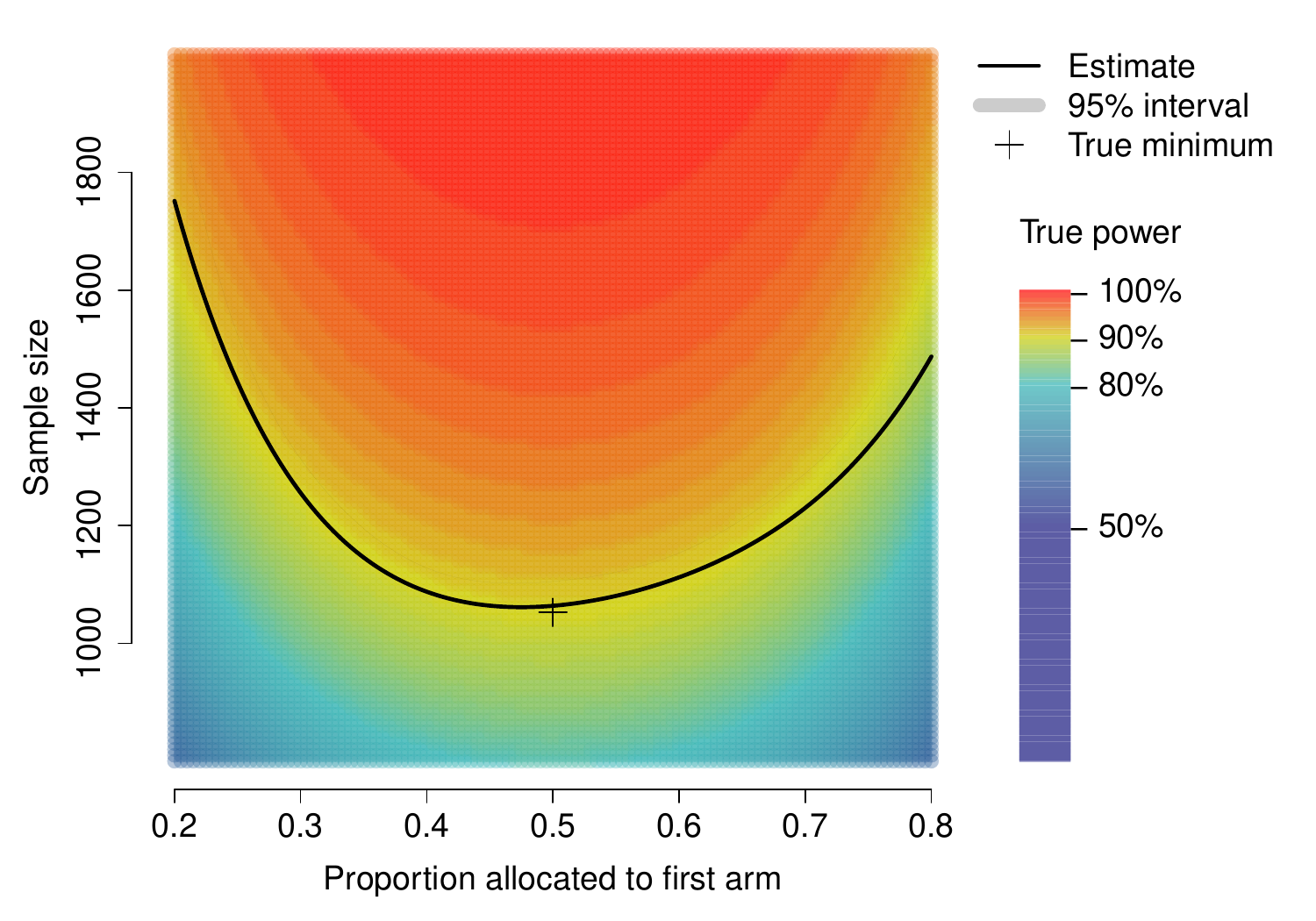}
\caption{\label{fig:oneacc} Empirical estimate of the relationship between allocation fraction for one arm and sample size to achieve 90\% power. Background colors show closed-form analytical power : colloquially speaking, \code{adsasi\_1d} tried to find the yellow range, and in particular the point marked by a + sign, which shows the analytical sample size for 90\% power at the optimal allocation fraction of 50\% (balanced arms).} 
\end{figure}

The other wrapper of the package, \code{adsasi_1d}, explores one design parameter from the user-defined simulation function, which the user must point to in the \code{adsasi_1d} function call. We use the same example as before, where \code{f1} represented the fraction allocated to the first arm, but this time it will be allowed to vary within \code{adsasi_1d}. The function itself does not need to be changed, it can just be called with 
\begin{CodeChunk}
\begin{CodeInput}
R> set.seed(0)
R> adsasi_1d(simulate_unequal_t_test, 
+    optivar = "f1", optiwin = c(.2, .8), 
+    nsims = 10000, capNN = 5000)
+    )
\end{CodeInput}
\end{CodeChunk}
Here we use the argument \code{optivar} to indicate which argument of our simulation function will be optimized for, and \code{optiwin} to indicate the window that will be explored. In this example we change the fraction of patients allocated to one arm and will try values from 0.2 to 0.8. We run more simulations to get a more precise estimation over different allocation fractions, and plot the resulting run in Figure~\ref{fig:oneacc} versus the closed-form power, which is known for this example. The purely empirical determination with \code{adsasi_1d} is very close to the true values. 

\renewcommand{\arraystretch}{1.4}
\begin{table}[htbp]
\centering
\begin{tabular}{lp{12cm}}
\hline
Argument           & (type) Usage \\ 
\hline
\code{simfun}            & (function) The user-supplied function that describes the clinical trial scenario (or similar experiment) that needs to be explored. Must have as named arguments a sample size (named \code{NN}) and an arbitrary number of design parameters. Must return a boolean indicating whether the trial is successful or not, after performing any required computations (regressions, bootstraps) as written by the user, and never return \code{NA}. \\
\code{tar\_power}            & (single number between 0 and 1) Target power (or more broadly, probability of success). \code{adsasi\_0d} will seek regions where \code{simfun} returns \code{TRUE} with a frequency of \code{tar\_power}  , assuming that higher sample size equals higher probability of success. \\
$\ldots$               & Additional named arguments to be passed to \code{simfun}. Some of these arguments can be functions themselves (e.g., for trying different analysis models). Any \code{simfun} argument without a default value must be specified here. \\
\code{nsims}               & (single number) Number of simulations to be run. After initialization, simulations are run in batches of 10\% of the number of existing simulations, until \code{nsims} is reached. \\
\code{verbose}               & (boolean) Whether to print extra diagnostics messages. \\
\code{impNN}               & (single number, or infinity) Sample size that is considered impossible (either computationnally, or logistically). The simulator will exit if, after 500+ simulations, it looks like the best value is above this. In practice, is mostly useful to avoid expensive computations in situations where simfun is not written well or is prohibitively long to compute for large sample sizes. \\
\code{capNN}               & (single number, or infinity) Maximum sample size that will be simulated. Values above \code{capNN} are reduced and jittered between 90\% and 100\% of \code{capNN}. Also mostly useful to avoid expensive computations. Values between \code{capNN} and \code{impNN} will be extrapolations of unclear validity, so if it looks like the answer is really above \code{capNN}, try running again with a higher \code{capNN}. \\
\code{initiation}               & (boolean, or numeric matrix) Either a boolean indicating whether or not to keep the first 150 simulations for the relationship inference (those tend to be far from \code{tar_power}), or a matrix with simulation results from a previous run which the user wants enrich with more simulations (formatted exactly as produced by \code{adsasi_0d} with the same \code{simfun} and \code{keepsims = TRUE}). \\
\code{savegraphs}               & (boolean or string) Whether to save graphs on drive (vs. showing them in the console). If string, is interpreted as a typical name to be used (several graphs will be drawn, with iteration number, timestamp and .png file extension appended). The string can contain a filepath, but folders must already exist (e.g., with \code{dir.create()} from \pkg{base}, if automated). \\
\code{keepsims}               & (boolean) Whether to keep the simulations sizes and individual simulation outcomes in the output.  \\ \hline
\end{tabular}
\caption{\label{tab:argzero} \code{adsasi\_0d} arguments.}
\end{table}

\section{Function arguments and standard outputs}\label{sec:args}

Up to now, we have made simple calls to our wrapper functions, omitting some arguments and argument names. To be more systematic, the user may want to refer to Table~\ref{tab:argzero} with  all arguments for \code{adsasi_0d}. Manipulating \code{initiation}, \code{savegraphs} and \code{keepsims} is useful to resume runs, save graphs or plot them in a custom manner, respectively, but is not required for the main use of \code{adsasi_0d}. 

\begin{figure}[t!]
\centering
\includegraphics{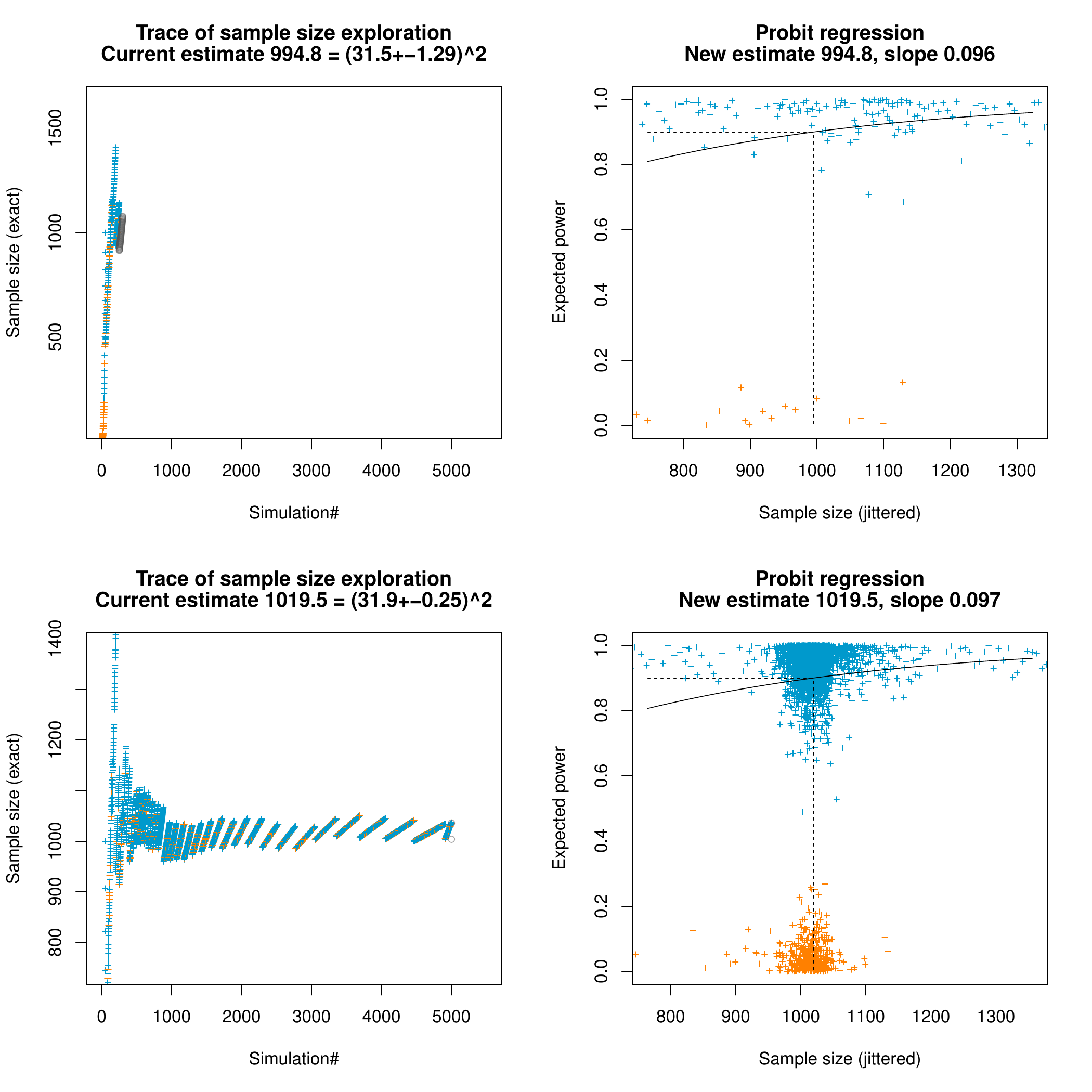}
\caption{\label{fig:zerorun} In-run diagnostics as shown by \code{adsasi\_0d} (top) after 250 simulations and (bottom) at the end of the run, using the first call from Section~\ref{sec:zeroacc}. Left panels show simulated sample sizes as a function of run progression (batches are visible as lines with consecutive simulations). Right panels show the regression, with successful simulations at the top (blue) and unsuccessful ones at the bottom (orange), with the modified probit regression line overlayed. Sample sizes are jittered on the right to better appreciate density.} 
\end{figure}

The most ergonomic way to use \code{adsasi_0d} is to simply look at the the in-console plots during a run, which are shown in Figure~\ref{fig:zerorun}. The user can follow each run and interrupt it if a problem seems evident (e.g., sample size too high due to an error in the user input function). The output object, unless \code{keepsims = TRUE}, is simply a list with one element, the estimated sample size, which is also shown on graphs. Automating several calls, however, will require saving outputs and potentially using \code{savegraphs = TRUE} or \code{keepsims = TRUE} to visualize the quality of the runs. 

For \code{adsasi_1d}, some additional arguments are used (some of them mandatory, some of them not), pertaining to the design parameter than needs to be probed, and to the flexibility of the model. The additional arguments for \code{adsasi_1d}, are shown in Table~\ref{tab:argone}, all arguments from \code{adsasi_0d} in Table~\ref{tab:argone} are also used for \code{adsasi_1d}. \code{optivar} (the design parameter of interest) is a mandatory argument and must match an argument from \code{simfun}. \code{optiwin} (the window to be explored) ranges from 0 to 1 by default, but this will often not make sense for whatever the user has written in \code{simfun}. 

\renewcommand{\arraystretch}{1.4}
\begin{table}[t!]
\centering
\begin{tabular}{lp{12cm}}
\hline
Argument           & (type) Usage \\ 
\hline
\code{optivar}            & (single string) Name of the \code{simfun} argument that needs to be optimized. \\
\code{optiwin}            & (numeric vector of size 2) Bounds of the region to be explored for values of \code{optivar}. \\
\code{optilog}               & (boolean) Whether \code{optivar} is best explored and drawn in log scale (as in the case of a ratio) or linearly. If for example \code{optiwin = c(0.1, 10)} for a ratio, graphs will be drawn with 1 as the middle value if \code{optilog = TRUE} and 5.05 if \code{optilog = FALSE}.. \\
\code{optiround}               & (boolean) Whether optivar needs to be rounded to the nearest integer to make sense for \code{simfun} (for example, if it is a number of centers in a cluster-randomized trial). \\
\code{n\_slope\_coefs}               & (single integer) Number of coefficients for the slope polynomial. The slope polynomial tries to model the relationship between \code{optivar} and the loss of power as sample size locally deviates from the (imperfectly known) target. \\
\code{n\_size\_coefs}               & (single integer) Number of coefficients for the size polynomial. The size polynomial tries to model the relationship between \code{optivar} and the target sample size. Its shape is the most useful output of the function. \\
\hline
\end{tabular}
\caption{\label{tab:argone} Additional arguments for \code{adsasi\_1d}.}
\end{table}

\begin{figure}[t]
\centering
\includegraphics[width=1\textwidth]{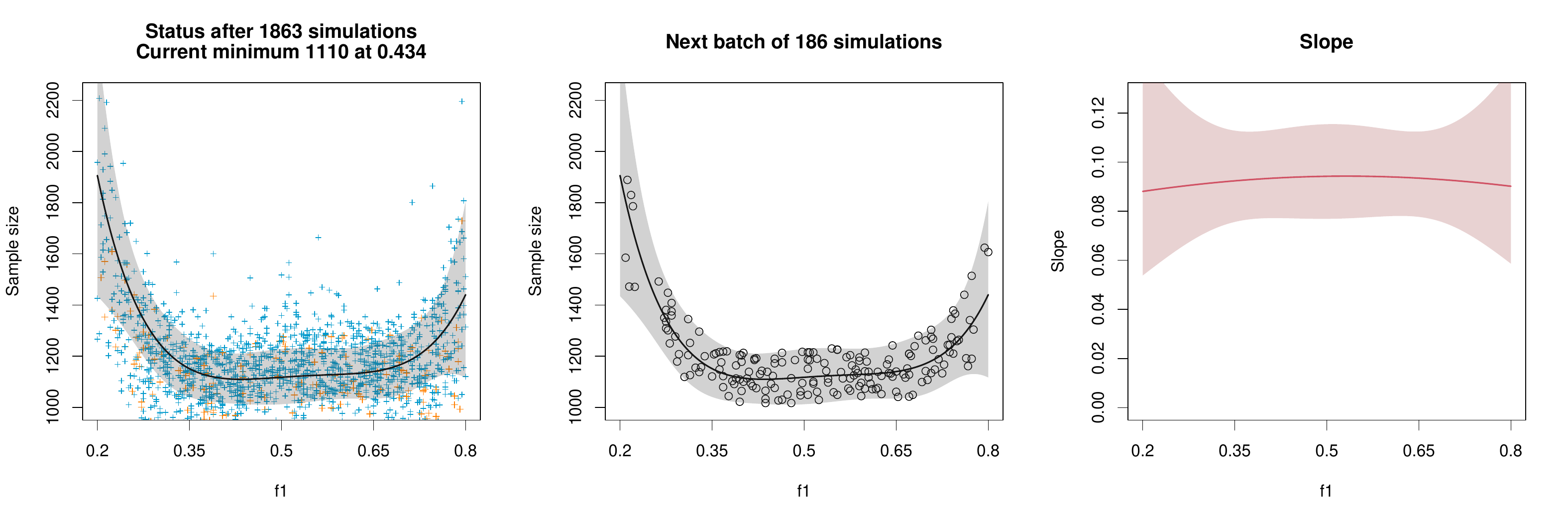}
\caption{\label{fig:onerun} In-run diagnostics as shown by \code{adsasi\_1d}. (left) Location of simulations on the design parameter $\times$ size surface, with successful ones in blue and unsuccessful ones in orange. The current estimate of the relationship between sample size and parameter of interest is overlayed. (middle) Location of the next batch of the simulations, with the same overlay. (right) Relationship between slope $e^{s\left(v\right)}$ and parameter of interest $v$, only useful, if at all, to verify regular behavior.} 
\end{figure}

Calling \code{adsasi_1d} produces graphs of the type represented in Figure~\ref{fig:onerun}, and iterate them throughout the run (the output is from out $t$~test example). The underlying data can be extracted by calling \code{adsasi_1d} with \code{keepsims = TRUE} and saving the output in an object. Assuming such a call has been saved in an object named \code{batch}, the following code can overlay the sample size estimates and their 95\% confidence intervals on an existing plot (something similar was used for Figure~\ref{fig:oneacc}). 
\begin{CodeChunk}
\begin{CodeInput}
R> lines(x = batch[["abscissae"]],
+        y = batch[["size_natural_estimate_by_optival"]]
+        )
R> polygon(x = c(batch[["abscissae"]],
+                rev(batch[["abscissae"]])),
+          y = c(batch[["size_confint_higher_by_optival"]],
+                rev(batch[["size_confint_lower_by_optival"]])),
+          col="#00000088",border=NA
+          )
\end{CodeInput}
\end{CodeChunk}

The number of simulations can be difficult to decide with \code{adsasi_1d} (it will depend on the degrees of freedom and level of misspecification of the polynomials), but generally 5000 to 10000 simulations are needed to get a robust shape of the relationship between the parameter of interest and sample size. 

\section{Intractable examples}\label{sec:intractables}
Having shown the algorithms' validity in a simple case with a closed-form solution (that can be used to validate against, but that also makes them not particularly useful), we now move to more complex cases where \pkg{adsasi} can substitute for non-existing closed-form solutions. 
\subsection{Ranking approved treatments} \label{sec:ranktrial}
We turn our attention to the problem of choosing between several available first-line treatments. This situation is encountered in indications ranging from hypertension to oncology. We assume that 5 treatments are available, and want our trial to show 2 things : first, using an omnibus test, it needs to observe a difference between the 5 treatments. Second, it needs to correctly identify the best 2. 

To generate the data, we assume a normal quantitative outcome and a difference of 0.5 standard deviation between the worst and the best treatment (so the means for the 5 groups are 0, 0.125, 0.25, 0.375 and 0.5, and our trial needs to correctly pick those with means 0.375 and 0.5 as the best). For analysis, we use rank-based statistics, namely the Kruskall-Wallis test followed by a ranking of treatments by the average outcome rank of their allocated patients. 

\begin{figure}[t!]
\centering
\includegraphics{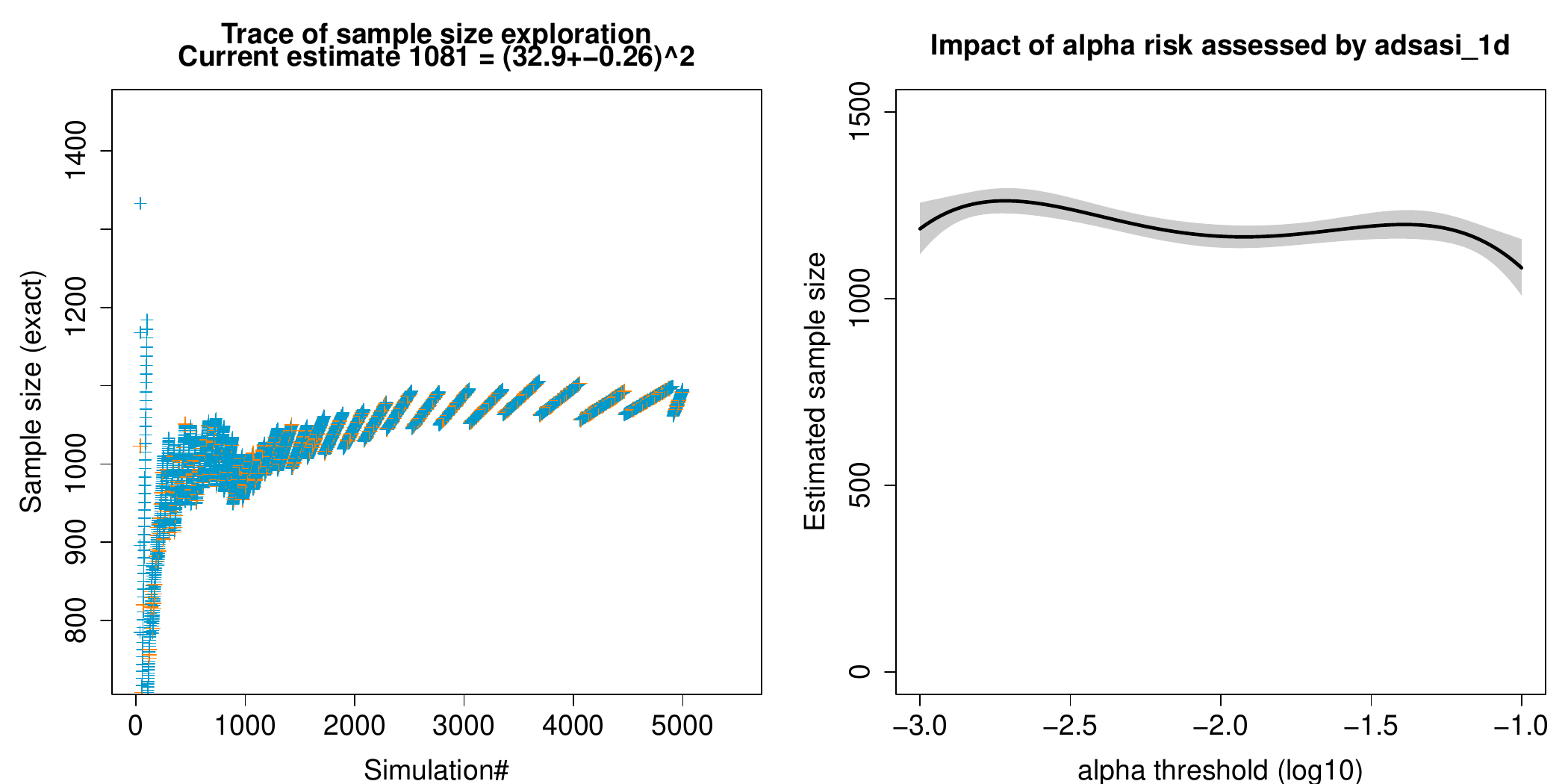}
\caption{\label{fig:figrank} Simulations for the ranking-based trial. (left) Partial graphical output from \code{adsasi\_0d}. (right) Impact of $\alpha$ risk of the omnibus Kruskall-Wallis test on sample size (partial output from \code{adsasi\_0d}).}
\end{figure}

Using a simulation approach, we can program the two success conditions in our simulation function, as shown below (\code{pp} and \code{rr}). 
\begin{CodeChunk}
\begin{CodeInput}
R> simulate_ranking = function(NN = 100, means = seq(0,.5, .125), alpha=.05)
+    { 
+     n_cand = length(means)
+     Ns = rep(round(NN/n_cand), n_cand - 1)
+     Ns = c(Ns, NN - sum(Ns))
+     datalist = lapply(1:n_cand,function(xx){rnorm(Ns[xx], means[xx])})
+     pp = NA
+     try(pp <- kruskal.test(datalist)$p.value)
+     dd = cbind(arm = rep(1:n_cand,Ns), y = do.call(c, datalist))
+     arm_mean_ranks = sapply(1:n_cand, function(xx){
+                        mean(rank(dd[, "y"])[dd[, "arm"] == xx])})
+     arm_ranks = rank(arm_mean_ranks)
+     rr = all(arm_ranks[(-1:0) + n_cand] > (n_cand - 1.5))
+     !is.na(pp) & pp < alpha & rr
+     }
\end{CodeInput}
\end{CodeChunk}
Without any other setup, we can then run \code{adsasi_0d} to find our sample size. 
\begin{CodeChunk}
\begin{CodeInput}
R> set.seed(0)
R> adsasi_0d(simulate_ranking, nsims = 5000, capNN = 10000)
\end{CodeInput}
\begin{CodeOutput}
$size_estimate
[1] 1081
\end{CodeOutput}
\end{CodeChunk}
This is in contrast with a parallel-arm trial which would need at least 6730 patients (as determined by a $t$~test closed-form calculation with effect size 0.125, scaled up to 5 arms). 
\begin{CodeChunk}
\begin{CodeInput}
power.t.test(NULL, delta = .125, sig = .05, power = .9)$n * 5
\end{CodeInput}
\begin{CodeOutput}
[1] 6729.556
\end{CodeOutput}
\end{CodeChunk}
As is often the case in clinical methodology, allowing more flexibility in the formulation of the clinical problem enables a drastic reduction in sample size. Finally, \code{adsasi_1d} allows the user to probe the impact of a design parameter. We apply this capability to the $\alpha$ risk of the omnibus Kruskall-Wallis test. 
\begin{CodeChunk}
\begin{CodeInput}
R> set.seed(0)
R> adsasi_1d(simulate_ranking, optivar = "alpha", optiwin = c(.001, .1), 
+     optilog = TRUE, nsims = 20000, capNN = 10000)
\end{CodeInput}
\end{CodeChunk}
The results from both \code{adsasi_0d} and \code{adsasi_1d} runs are summarized in Figure~\ref{fig:figrank}. The \code{adsasi_1d} output shows that the $\alpha$ risk of the omnibus Kruskall-Wallis test has little impact on sample size. 

\subsection{Sampling from observed data} \label{sec:bootstrap}
All examples so far generate the data from scratch using relatively simple random draws. Being based on simulations, \pkg{adsasi} can also accommodate much more complex data structures with hard-to-describe internal correlations, chief among them data from existing patients. This is especially useful when the planned analysis is a multivariate model that may depend on difficult-to-describe conditional correlations. 

For this example, we use data from the International Stroke Trial (IST)  \citep{international_stroke_trial_collaborative_group_international_1997}. An easily imported version of the dataset can be found in this article's replication materials, adapted from a public repository entry \citep{sandercock_international_2011}. IST followed stroke patients and collected their concomitant treatments as well as in-hospital mortality, and randomized patients into receiving anticoagulants or not. We ignore anticoagulants and focus instead on calcium antagonists, one of the non-randomized treatments that was also recorded (hence requiring adjustment). We restrict the analysis to patients with ischemic stroke, and to the outcome of in-hospital death (N=17389 patients, among which n=9355 survivors, in this 1997 trial). 

\begin{figure}[t!]
\centering
\includegraphics[width=0.9\textwidth]{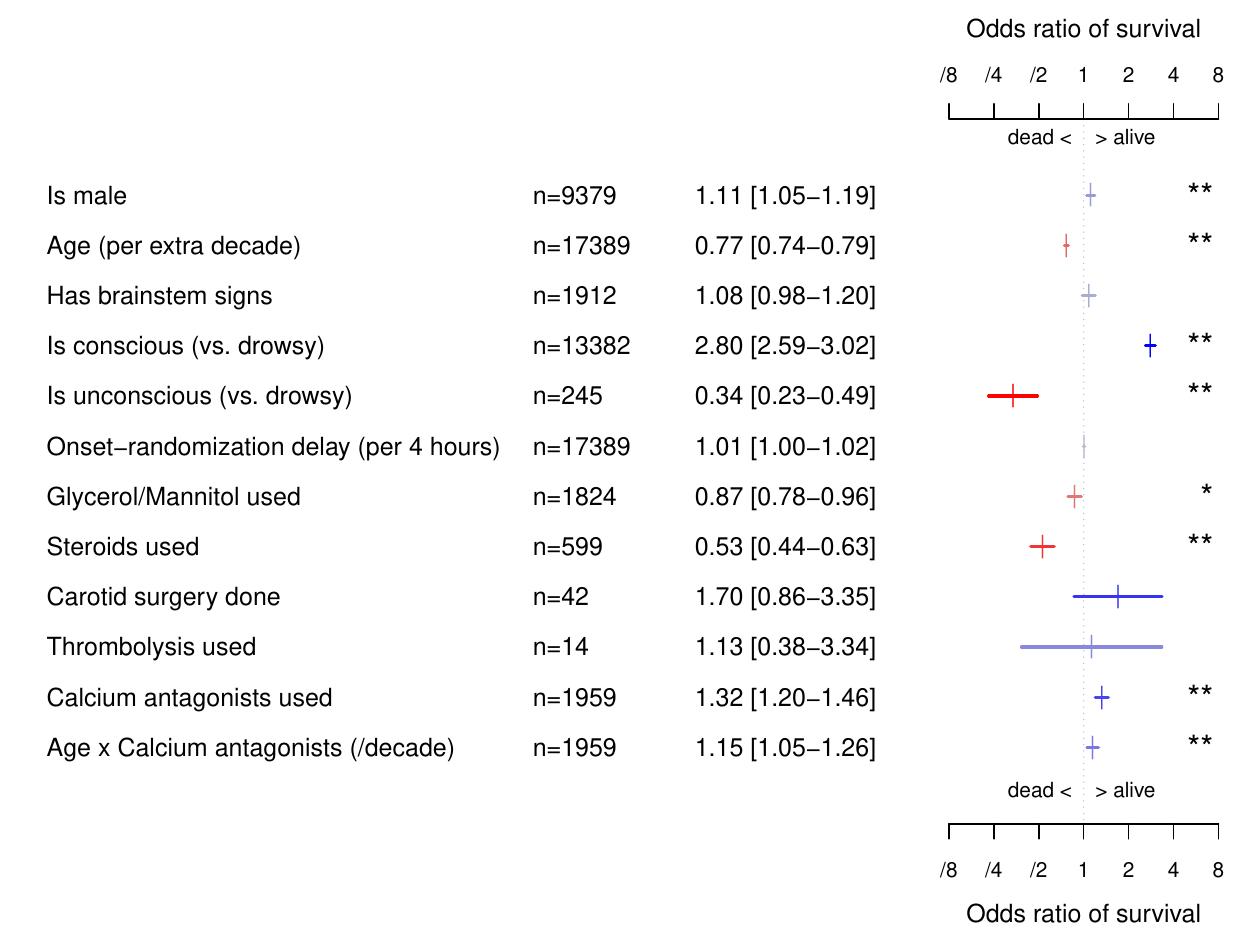}
\caption{\label{fig:forestplot} Observational logistic model fitted on the IST cohort. The $n$ figures are for the number of informative patients (that either have the binary variable or have a value for the continuous variable). The shown dataset is with imputation at the mean (<2\% cases) except for procedures (for which missing was interpreted as not done), and total cohort size is 17389. $*$ denotes $p<0.05$, $**$ denotes $p<0.005$ (unadjusted for multiplicity).}
\end{figure}

We first model death as a function of classical prognostic factors in stroke, namely sex, age, brainstem lesion signs, consciousness at admission, delay from onset of symptoms, and the recorded co-interventions in the trial : glycerol/mannitol, steroids, carotid surgery, thrombolysis and calcium antagonists. We find a significant effect of calcium antagonists and an interaction with age in a logistic model shown in Figure~\ref{fig:forestplot}. 

We then sample from this population to simulate a 1:1 randomized trial of calcium antagonists in this population. We assume that 75\% of the observed effect is due to the drug itself (25\% being residual confounding), and apply a counterfactual reasoning : if the simulated patient has their treatment switched compared to their counterpart in the original dataset, we apply a shift in their log odds of survival (the linear predictor in logistic regression) that is equal to 75\% of the distance between the predictions with and without calcium antagonists, using the model from the original dataset. We then draw all survival outcomes (shifted or not), and model the simulated dataset using a logistic regression adjusted for sex and age. Whether or not the calcium blocker coefficient is significant gives us the result of our trial. We call the function \code{bootstrap_ist} and do not show it for brevity (is is provided in the replication materials). 
\iffalse
\begin{CodeChunk}
\begin{CodeInput}
R> bootstrap_ist = function(NN, min_age = 0, residual = .25, alpha =.05)
+    {
+     ss = sample((1:nrow(ist))[ist[, "AGE"] > ((min_age - mean_age) / 10)],
+                   NN, replace=TRUE) 
+     picks = ist[ss, vars]
+     picks[, "DCAA_old"] = picks[, "DCAA"]
+     picks[, "DCAA_new"] = rbinom(NN, 1, .5)
+     logodds_default = predict(model, newdata = picks, type = "link")
+     picks[, "DCAA"] = 0
+     logodds_all_control = predict(model, newdata = picks, type = "link")
+     picks[, "DCAA"] = 1
+     logodds_all_active = predict(model, newdata = picks, type = "link")
+     logodds_alive = logodds_default +
+                       (picks[, "DCAA_new"] - picks[, "DCAA_old"]) *
+                       (logodds_all_active - logodds_all_control) * (1 - residual)
+     picks[, "DCAA"] = picks[, "DCAA_new"]
+     picks[, "DALIVE"] = runif(NN) < (1 / (1 + exp(-logodds_alive)))
+     pp = NA
+     try({
+          model = glm("DALIVE ~ SEX + AGE + DCAA", data = picks, family = binomial)
+          pp <- summary(model)$coef["DCAA", "Pr(>|z|)"]
+          })
+     !is.na(pp) & pp < alpha
+     }
\end{CodeInput}
\end{CodeChunk}
\fi

\begin{figure}[t!]
\centering
\includegraphics{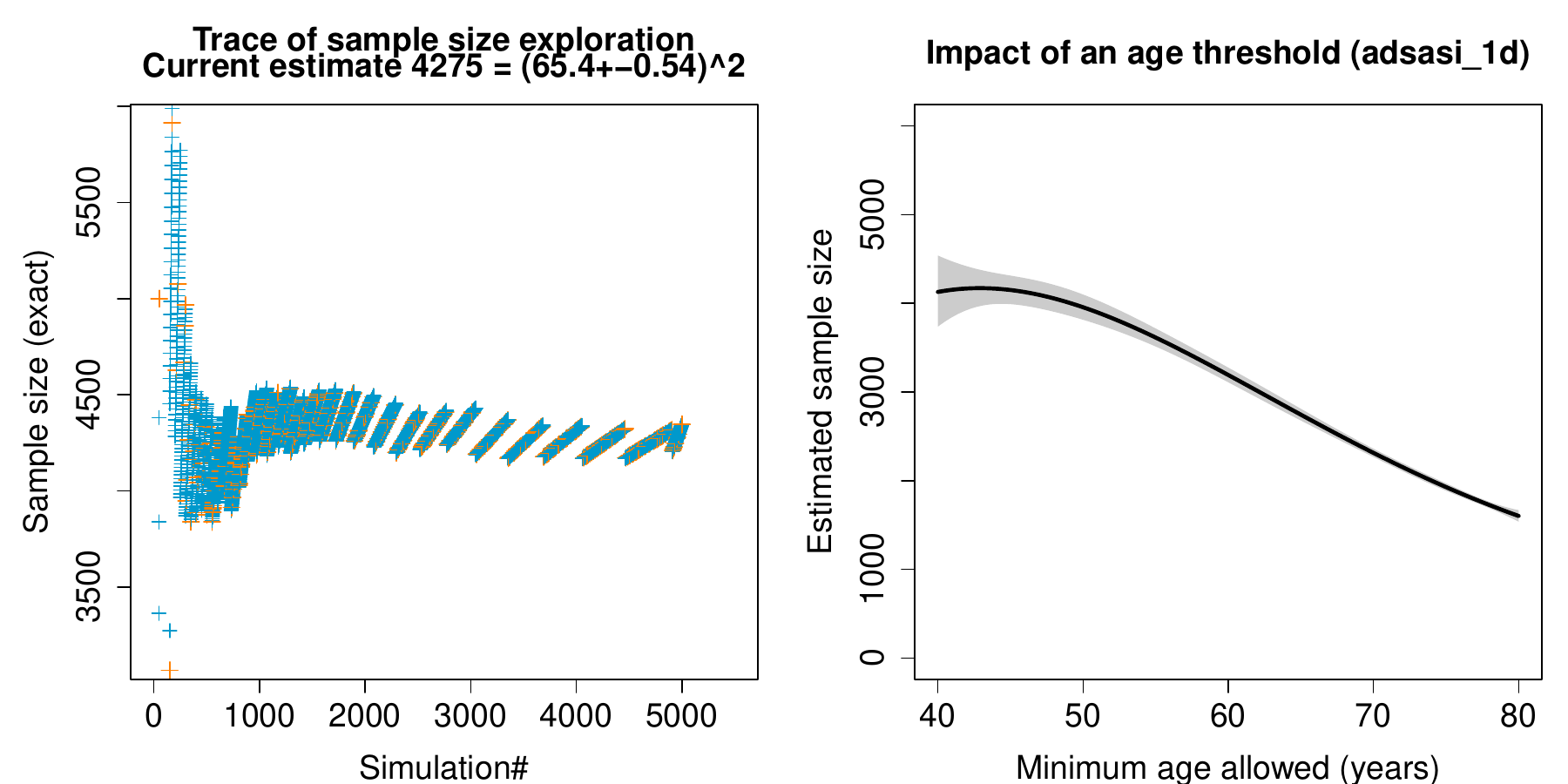}
\caption{\label{fig:bootstrap} Simulations from an existing patient cohort. (left) Partial graphical output from \code{adsasi\_0d}. (right) Impact of minimum age inclusion threshold on sample size (partial output from \code{adsasi\_0d}).}
\end{figure}

Once \code{bootstrap_ist} is written, calls to \code{adsasi_0d} and \code{adsasi_1d} are straightforward. Because of the apparent interaction with age, one legitimate question is to limit inclusion to more elderly patients in order to increase power, which we can answer with an \code{adsasi_1d} call. 
\begin{CodeChunk}
\begin{CodeInput}
R> set.seed(0)
R> adsasi_0d(bootstrap_ist, nsims = 5000, capNN = 10000)
R> set.seed(0)
R> adsasi_1d(bootstrap_ist, optivar = "min_age", optiwin = c(40, 80), 
+     nsims = 20000, capNN = 10000)
\end{CodeInput}
\end{CodeChunk}
The results are shown in Figure~\ref{fig:bootstrap}. Recruiting older patients does indeed increase power significantly (right panel), down to around 1500 patients with elderly patients only (>80 years old). The estimation is less precise for lower age thresholds because the simulation parameter sampling is done preferentially around better regions of the design parameter, which increases precision locally. 

Of note, a recent Cochrane review found no evidence for calcium channel blockers in acute stroke, although elderly patients were not a subgroup of interest \citep{zhang_calcium_2019}, highlighting limits of extrapolation from observational data. Nonetheless, observational data will be the only one available when planning the first randomized trial on a given question. 
\section{Discussion} \label{sec:discussion}
\subsection{Best use cases} \label{sec:usecases}
The philosophy of \pkg{adsasi} is to be as broad as possible : any simulation that the user can write, \pkg{adsasi} can be used to find a sample size for. Users that will benefit most are experimental methodologists that are comfortable with writing simulation functions from scratch (of the kind shown in this article). The goal is to enable them to devise highly original approaches for which no closed-form sample sizes can be computed. Any particular new approach that becomes particularly popular is likely to have a closed form sample size approximation derived at some point, so it may be the case that \pkg{adsasi} will mostly be useful in the discovery and ideation phase. But before that popularization, and in the back-and-forth discussions between experimenters and methodologists, having a tool like \pkg{adsasi} allows the latter more creativity. 

Another use case that we have observed is when a sample size formula exists, but includes parameters that are ill-defined, difficult to understand or imagine. Clinical data often does not come in the exact form that a formula expects, and while it will generally be possible to derive the appropriate parameters by simulation, \pkg{adsasi} can be an interesting check on that computation (once the simulation code is written, running \pkg{adsasi} functions is trivial anyway). For example, in the context of cluster-randomized trials, differences between rates of favorable outcomes between different hospitals can be obtained, but intra-class correlation coefficients for logistic models require some work. 

Finally, \pkg{adsasi} will be useful whenever simulations are run for another reason (e.g., to check type I error), since the simulation function can be passed without any modification to the \pkg{adsasi} wrappers. Clinical trial simulation is increasingly frequent for reasons not strictly related to sample size, and in those cases a quick detour through \pkg{adsasi} may provide peace of mind to the experiment designer. 

\subsection{Limits and future directions} \label{sec:usecases}
The main limit of \pkg{adsasi} is that it runs the simulation a large number of times near the final sample size (generally a few thousand, depending on the desired accuracy). Thus, simulations with slow inference such as Markov chain Monte Carlo (MCMC) Bayesian trials, or stochastically activated expectation maximization (SAEM) for non-linear mixed effect models, although feasible, will not be convenient use cases. Some simpler models also have relatively slow inference, such as win ratio regressions \citep{duarte_winratio_2020}. 

On a related note, the setup of parallel cores was found to slow down the execution in simple cases, so the option was not added by default, but future versions will likely add it as an option, which may be worth it for slower inference procedures (advanced users may also be able to do it themselves by changing the single line that calls the simulation function within \code{adsasi_0d} and \code{adsasi_1d}, which are written in base \proglang{R}). 

Another limit is that \pkg{adsasi} requires the user to write very robust simulation functions, of the kind we have shown here (e.g., with liberal use of \code{!is.na()\&} and \code{try()}). Since a large number of stochastic simulations will be run by \pkg{adsasi} functions at wildly different sample sizes, including small ones, many failures are possible (perfect separation, insufficient number of random effect levels, no patients in one arm, etc.). Unless the user is systematic in their use of robust commands like \code{!is.na()\&} and \code{try()}, using \pkg{adsasi} can sometimes be frustrating. Some functions or packages are not really be meant to be run automatically a large number of times (e.g., complex models with different optimization parameters to choose from by trial and error, such as in \pkg{lme4} \citep{bates_lme4_2026}), and they require some practice with error handling to be fit into an \pkg{adsasi} workflow. 

Finally, the user may want to optimize more than one parameter simultaneously. Those can currently only be done sequentially with \code{adsasi_1d}. However, in practice, it is not that common to have absolutely no constraint on, and no idea about, many parameters in a design. Still, introducing high-dimensional polynomials should be feasible, although significantly heavier than what is currently implemented in \code{adsasi_1d}, and might be taxing in computer time. For this, implementing the Hessian computation more efficiently is also a promising future direction. 

\section{Summary and conclusion} \label{sec:conclusion}
In this paper, we present, analyze and familiarize the reader with the \proglang{R} package \pkg{adsasi}. The \code{adsasi_0d} function can empirically find the sample size for any clinical trial whose simulation can be written and executed relatively quickly, by sharing information across different tested sample sizes seamlessly. 

The other function in the package, \code{adsasi_1d}, can optimize for an arbitrary design parameter in a given window. The information sharing is also done across the design parameter dimension, using an arbitrary polynomial whose unavoidable mis-specification is mitigated by sampling simulated sizes close to the true sample sizes. 

The two functions were first used on an example with a closed-form solution to check their accuracy, and then showcased in more complex ones. First, a ranking-based outcome of a trial was examined and dimensioned, showing a drastic reduction over a similar design with pair-wise comparisons. Second, a resampling-based approach was shown, where an existing cohort of patients provided a dataset with complex correlation structures, on which complicated counterfactual reasonings and adjusted analyses could be applied without sacrificing precise sample size computations. 

%% -- Optional special unnumbered sections -------------------------------------

\section*{Computational details}

Simulations were run with \proglang{R}~4.2.2. for Windows (Microsoft Corporation, Redmond, Virginia) on a single i7 Gen 12 core (i7-1265U, Intel corporation, Santa Clara, California). Figures were drawn with base \proglang{R} and built-in packages \pkg{grDevices}, \pkg{graphics}, and \pkg{stats} using the code provided in the \pkg{adsasi} functions help. \pkg{adsasi}~0.9.0.2 and its base \proglang{R} dependency \pkg{abind} were used, with some custom modifications of the functions to generate vector-based graphics. \pkg{readxl} was used to load a file and \pkg{qpdf} was used to rearrange figures, but neither is used by \pkg{adsasi} itself. Figure data were generated with a random seed set at 0 before running the \pkg{adsasi} functions, every time. All code is provided as run by the author to replicate all figures, and runs in about 1 hour on the aforementioned machine. 

\proglang{R} itself and all packages used, including \pkg{adsasi}, are available from the Comprehensive \proglang{R} Archive Network (CRAN) at \url{https://CRAN.R-project.org/}.

\section*{Acknowledgments}

No specific funding was used for this work. The author would like to thank his parent institutions for the academic freedom they provide, which has often been much more fruitful than grants. Colleagues from DEBRC and IAME, especially Romain Leroux and France Mentré, also deserve special thanks for helpful discussions and encouragements. 

%% -- Bibliography -------------------------------------------------------------
%% - References need to be provided in a .bib BibTeX database.
%% - All references should be made with \cite, \citet, \citep, \citealp etc.
%%   (and never hard-coded). See the FAQ for details.
%% - JSS-specific markup (\proglang, \pkg, \code) should be used in the .bib.
%% - Titles in the .bib should be in title case.
%% - DOIs should be included where available.

\bibliography{refs}

%% -----------------------------------------------------------------------------

\end{document}